\documentclass[12pt]{article}
\usepackage{amsmath}
\usepackage{amssymb}
\usepackage{amsfonts}
\usepackage{latexsym}
\usepackage{graphicx}
\usepackage{latexsym}
\usepackage{color}

\newcommand{\eq}[1]{(\ref{#1})}

\newcommand{\n}[1]{\label{#1}}

\def\beq{\begin{eqnarray}}
\def\eeq{\end{eqnarray}}
\def\ln{\,\mbox{ln}\,}

\def\Det{\,\mbox{Det}\,}
\def\det{\,\mbox{det}\,}
\def\tr{\,\mbox{tr}\,}

\def\Tr{\,\mbox{Tr}\,}

\def\al{\alpha}
\def\be{\beta}

\def\de{\delta}
\def\vp{\varepsilon}

\def\la{\lambda}
\def\na{\nabla}
\def\pa{\partial}

\def\om{\omega}
\def\ph{\varphi}

\def\Ga{\Gamma}

\usepackage{indentfirst} 

\begin{document}

\begin{center}

{\Large
Massive vector field on curved background: non-minimal
coupling, quantization and divergences
}

\vskip 6mm


\textbf{Ioseph L. Buchbinder}$^a$ \footnote{ E-mail:
josephb@tspu.edu.ru}, \textbf{Tib\'{e}rio de Paula Netto}$^b$
\footnote{ E-mail: tiberiop@fisica.ufjf.br}, \textbf{Ilya L.
Shapiro}$^{b,a}$ \footnote{ E-mail: shapiro@fisica.ufjf.br} \vskip
4mm

(a) Department of Theoretical Physics, Tomsk State Pedagogical
University, \\
634061, Tomsk, Russia; \\
National Research Tomsk State University, 634050, Tomsk, Russia\\

(b) \ Departamento de F\'{\i}sica, \ ICE, \
Universidade Federal de Juiz de Fora
\\ Juiz de Fora, \ 36036-330, \ MG, \ Brazil

\end{center}
\vskip 12mm

\begin{quotation}
\noindent
{\large {\it Abstract}}.
\quad
We study the effective action for the massive vector field theory
non-minimally coupled to external gravitational field. Such a theory
is an interesting model both from the theoretical side and also due
to the various phenomenological applications to cosmology and
astrophysics. The present work pretends to initiate a systematic
study of its properties at the quantum level, by exploring free
massive vector coupled to an external symmetric second-rank tensor.
Stueckelberg scalar field is used to restore the gauge invariance.
After that, by using a special gauge fixing and non-local in
external fields change of variables, we diagonalize the bilinear
form of the action and develop a consistent procedure to study the
effective action. As a result we derive a complete non-linear
structure of divergences of the effective action and discuss its
properties.

\vskip 3mm

{\it MSC:} \
81T15,    
81T20     
\vskip 3mm

{\it PACS:} \
04.62.+v, 
11.10.Gh  
04.50.Kd  

\end{quotation}
\vskip 4mm

\section{Introduction}
\label{int}

The vector field models with soft and/or spontaneous violation of
gauge invariance is one of objects of modern study in gravity and
cosmology (see, e.g., \cite{PP,Dim1} and references therein). These
models are closely related to the classical and quantum vector
Galileons
\cite{Tasinato,Heis-2014,DeRham-2014,HeisKimYam,Farid,LH-mn}.
The list of applications of such models includes, in particular, the
generation of the initial seeds of magnetic fields in the early
universe \cite{WT} and vector inflation \cite{vec-in} (see also
\cite{Jabbari,Uzan} for extensive reviews). Both these subjects
attract a great deal of interest, however the considerations are
concentrated mainly on the classical theory. It seems natural to
extend the study of these theories to the quantum level, especially
because we know that quantum effects play an important role in the
usual scalar inflaton models. Therefore we can hope that the quantum
effects can play some role in the vector inflation.

In the present work we consider quantum aspects of massive Abelian
vector field in curved space-time. The first observation is that the
coupling of such a field to external gravity includes some
interesting non-trivial aspects. Indeed, let us start from an
arbitrary matter field model in Minkowski space. As we know, a
generalization of  such a model to curved space-time in non-unique
(see, e.g., \cite{book}).  The first possibility is to apply the
procedure of  minimal covariant generalization, including
interaction with gravity. The last means one should treat the
field  as a tensor or spinor in curved space-time, replace partial
derivatives by the corresponding covariant derivatives, without
changing the canonical dimension of the field and trying to
preserve as much as possible global and gauge symmetries of
the initial flat-space theory.
However, after that there is still a freedom to enrich the
Lagrangian by arbitrary local terms containing the matter
field and the powers  of the curvature tensor with some
coupling constants. All these terms describe the non-minimal
coupling of matter field to gravity. If we also demand
that the matter sector does not contain the new scales in comparison
with theory in flat space, then all non-minimal coupling constants
must be dimensionless.

Taking into account the above arguments, it is evident that the only
possible non-minimal coupling for scalar field ${\phi}$ is
${\xi}R{\phi}^2$ where $R$ is a scalar curvature and ${\xi}$ is a
dimensionless constant of non-minimal coupling. For the massless
vector field, non-minimal interactions with only dimensionless
coupling constants is forbidden by gauge invariance. Furthermore,
for a fermionic  spin-$1/2$ field the non-minimal interaction to
gravity is ruled out by the canonical dimension of the field. Let us
now consider the massive vector field model. This model is not
gauge invariant in flat space, therefore the arguments based on
gauge invariance do not work anymore. As a result, the most
general action for massive vector field in curved space-time
without dimensionful coupling constants, has the form
\beq
S &=& \int d^4x \, \sqrt{-g} \, \left\{ - \frac{1}{4}\,
F_{\mu\nu}^2 - \frac{1}{2}\, m^2 A_\mu^2 - \frac{1}{2}\,
X^{\mu\nu}(x) A_\mu A_\nu \right\}\,.
\label{action}
\eeq
where
\beq
X_{\mu\nu}(x) &=&\,{\xi}_1 R g_{\mu\nu} \,+\,{\xi}_2 R_{\mu\nu}.
\label{mag}
\eeq
Here ${\xi}_1$ and ${\xi}_2$ are two dimensionless
non-minimal coupling constants. All other notations are standard.
This is the only admissible non-minimal extension for theory of free
massive vector field in curved space-time.

The versions of (\ref{action}) which are mostly used for existing
applications, are \beq X_{\mu\nu} &=& {\xi}_1 R g_{\mu\nu}\,, \quad
m \neq 0 \label{inf} \eeq for vector inflation \cite{vec-in} and
\beq X_{\mu\nu} &=& {\xi}_1 R g_{\mu\nu} \,+\,{\xi}_2 R_{\mu\nu}\,,
\quad m=0 \label{mag1} \eeq in case of generating magnetic fields
\cite{WT}. In this case the curvature-dependent terms are introduced
at the phenomenological level to break the local conformal invariance
of electromagnetic field. The coefficients ${\xi}_1$ and ${\xi}_2$
are chosen in each particular case from the phenomenological
arguments.

In more general theories there may be other possibilities for the
field $\,X_{\mu\nu}$, besides (\ref{mag}). For example, there
can be the an external field $X^{\mu\nu}$ which is not related to
curvature, but can be useful for the phenomenological reasons.
Another possibility originates if we consider a self-interacting
massive vector field model or massive vector field coupled to
other dynamical fields like scalars and/or fermions. Then in the
one-loop approximation one can meet the action (\ref{action}),
where the field $X^{\mu\nu}$ is constructed from background
vector, scalar or spinor fields. These circumstances motivate
us to consider the model with  action (\ref{action}), where
no concrete form for the field $X^{\mu\nu}$ is assumed.

In the present work we intend to study the model (\ref{action})
treating $X^{\mu\nu}(x)$ as an arbitrary external symmetric tensor
field. Our main purpose is to formulate the quantum effective
action $\,\Ga[g_{\mu\nu}, X^{\mu\nu}]\,$ for this model  and to
calculate its divergences. It is important to note that the naive
quantization of the model (\ref{action}) does not work. Within
the standard scheme the effective action should be defined as a
functional integral over the fields $A_{\mu}(x)$ of the exponential
of the action (\ref{action}). Hence the result would have the form
of the functional determinant of the operator related to the
quadratic part (bilinear form) of the action (\ref{action}). The
reason why this approach is not appropriate is that the kinetic
part of the action (\ref{action}) is gauge invariant while the
mass and non-minimal terms are non-invariant. This yields the
constraints on the dynamics of the theory and as a result the naive
functional integral approach becomes incorrect.

The first study of the model (\ref{action}) has been undertaken by
D.J. Toms \cite{Toms}. It was shown that the naive approach treating
the term with $\,X^{\mu\nu}\,$ as a small perturbation leads to
inconsistencies, because the propagator of the zero-order theory,
that is the Proca model, is a subject of the mentioned constraint.
This observation was confirmed by the canonical analysis.
In particular, this means that the standard procedure of dealing
with a free Proca model in curved space \cite{bavi85} (see also
\cite{fervi}) leads to an inconsistency\footnote{Certain aspects of
quantum Proca theory were recently discussed in \cite{BF}.}.
The derivation of the
one-loop divergences in \cite{Toms} was based on the non-covariant
approach related to the Faddeev-Jackiw quantization \cite{FJ}. As a
result the divergences were obtained only for the two special cases
with constant external field $X^{\mu\nu}$, from what it is difficult
to restore the covariant result.

In what follows we describe the general procedure of consistent
quantization of the theory (\ref{action}) which is manifestly
covariant and can be applied for an unconstrained and possibly
non-constant external field $\,X_{\mu\nu}(x)$ and for an arbitrary
external metric. As a result we arrive at the covariant expression
for the effective action and calculate its divergent part. As it was
already mentioned above, the quantization is not a trivial task
because of the degeneracy of the kinetic term and the broken
gauge symmetry. In the usual Proca theory this problem has two
standard solutions \cite{Stueckelberg}, but both methods meet
serious difficulties in the presence of the non-minimal term with
an arbitrary field $\,X^{\mu\nu}$. In the rest of the paper we
perform calculations by combining a non-standard application
of the covariant Stueckelberg procedure and different versions
of the Schwinger-DeWitt technique \cite{dewitt,bavi85}.
The work can be seen as a first step in exploring the quantum
properties and loop corrections for a wider class of massive
and massless vector theories with a non-minimal interaction

The paper is organized as follows. In Sect.~\ref{bs} it is shown how
the theory of massive vector field with  non-minimal coupling to
external tensor field can be reformulated by introducing
compensating auxiliary scalar and restoring gauge invariance. After
that the gauge fixing is introduced and the basis for quantum theory
formulated. In Sect.~\ref{bi} we construct the bilinear form of the
action and discuss two alternative ways to make it diagonal.
Sect.~\ref{qua} is devoted to the non-polynomial change of variables
in the scalar sector and to the main quantum calculation. In
Sect.~\ref{pert} an independent calculation the first order in
$X^{\mu\nu}$ is performed to provide an independent verification of
the main result. Finally, in Sect.~\ref{con} we draw our conclusions
and discuss possible extensions and generalizations of this work.

\section{Restoring and fixing gauge symmetry}
\label{bs}

Our consideration is based on reformulation of the model
(\ref{action}) in dynamical equivalent but gauge invariant form. For
this aim we use the Stueckelberg procedure \cite{Stueckelberg} and
perform the transformation
\beq
A_\mu \to A_\mu - \frac{1}{m} \,
\pa_\mu \ph \,. \label{St1}
\eeq
After that the new action, depending
now on $A_{\mu}$ and $\ph$, becomes invariant under the gauge
transformation
\beq
A_\mu \to A_\mu + \pa_\mu f \,,\qquad \ph \to
\ph + m f\,. \label{St2}
\eeq
Here
 $f$ is gauge parameter.
As a result, we arrive at the action which is dynamically equivalent
to (\ref{action}), but possesses the symmetry under (\ref{St2}),
\beq
S' &=& \int d^4x \, \sqrt{-g} \, \Bigg\{ - \frac{1}{4}\,
F_{\mu\nu}^2 - \frac{1}{2}\, m^2 A_\mu^2 - \frac{1}{2}\, X^{\mu\nu}
A_\mu A_\nu \label{Stuk}
\\
&-&
 \frac12 (\na \ph)^2
 - \frac{1}{2 m^2}\, X^{\mu\nu} \na_\nu \ph \, \na_\mu \ph
\,+\, m \, A^\mu \na_\mu \ph \,+\, \frac{1}{m} X^{\mu\nu} \, A_\nu
\, \na_\mu \ph \Bigg\}\,. \label{St3}
\eeq
In the special gauge
$\ph=0$ the theory comes back to (\ref{action}). According to the
standard consideration (see e.g. \cite{BuGui}) the result of the
quantum calculation in this and similar cases does not depend on the
gauge fixing condition, and hence one is free to choose  another
gauge, making quantum theory more transparent and calculations more
accessible.

A useful gauge fixing (GF) action is given by
\beq
S_{gf}
&=&
-\,\frac12 \int d^4x \, \sqrt{-g} \, \chi^2\,,
\label{St4}
\eeq
where
\beq
\n{gf1}
\chi = \na_\mu A^\mu - m \ph \,.
\eeq
The total action with the gauge fixing term has the form
\beq
S' + S_{gf}
&=&
\int d^4x \, \sqrt{-g} \, \Bigg\{
- \frac{1}{4}\, F_{\mu\nu}^2 -\frac{1}{2} \, (\na_\mu A^\mu)^2
- \frac{1}{2}\, m^2 A_\mu^2
- \frac{1}{2}\, X^{\mu\nu} A_\mu A_\nu
\nonumber
\\
&-& \frac{1}{2} (\na \ph)^2 - \frac{1}{2}\, m^2 \ph^2 -
\frac{1}{2m^2}\, X^{\mu\nu} \, \na_\mu \ph \, \na_\nu \ph +
\frac{1}{m} \, X^{\mu\nu} \, A_\nu \, \na_\mu \ph \Bigg\}\,,
\label{a1}
\eeq
which still looks complicated, especially due to the non-minimal
second-derivative term in the scalar sector. There is also a linear
in derivative term in the mixed $A\ph$-sector. Due to the unusual
scalar operator one can not apply a known technique (see e.g.
\cite{book}) of dealing with such a term and it is necessary to look
for some other approach.

\section{Bilinear operator in quantum fields}
\label{bi}

Using the GF condition \eq{gf1} and the action \eq{a1}
one can find the following operator of the theory:
\beq
S' + S_{gf}
&=&
\int d^ 4x \sqrt{-g} \big\{ {\cal L}' + {\cal L}_{gf}\big\}
\nonumber
\\
&=&
\frac{1}{2} \int d^4x \, \sqrt{-g}
\, \left(\begin{array}{cc}
A_\mu & \ph \end{array} \right)
\hat{\mathbf H}
\left(\begin{array}{cc}
A_\nu
\\
\ph \end{array} \right)\,,
\label{bim}
\eeq
where
\beq
\n{bili}
\hat{\mathbf H}
&=&
\left(\begin{array}{cc}
\hat{H}_{AA} & \hat{H}_{A\ph} \\
\hat{H}_{\ph A} & \hat{H}_{\ph \ph}
\end{array} \right)\,.
\eeq
Here and in the following we use bold notations for the matrix
operators only.  The blocks of (\ref{bili}) have the form
\beq
\hat{H}_{AA}
 &=&
g^{\mu\nu} (\Box - m^2) - R^{\mu\nu}
- X^{\mu\nu}\,,
\label{AA}
\\
\hat{H}_{A\ph}
 &=&
\frac{1}{m} \, X^{\mu\nu} \na_\mu\,,
\label{Ap}
\\
\hat{H}_{\ph A}
&=&
-\, \frac{1}{m} \, (\na_\mu X^{\mu\nu})
- \frac{1}{m} \, X^{\mu\nu} \na_\mu\,,
\label{pA}
\\
\hat{H}_{\ph\ph}
 &=&
\Box - m^2 + \frac{1}{m^2} \, (\na_\mu X^{\mu\nu}) \na_\nu +
\frac{1}{m^2} \, X^{\mu\nu} \na_\mu \na_\nu\,. \label{pp} \eeq We
emphasize here that no one term in the action (\ref{bim}) is
degenerated. It means that the quantization is realized by standard
procedure, the corresponding effective action is given by standard
functional integral for gauge theory and the effective action will
be given by \beq \frac{i}{2}\,\ln\Det \hat{\mathbf H} \label{det}.
\eeq  Plus the corresponding ghost contribution. However, the
bilinear form (\ref{bili}) has non standard structure for study the
effective action. For instance the operator $\hat{\mathbf H}$ is
non-minimal and non-diagonal at the same time. Our next purpose will
be to derive the divergent part of the expression (\ref{det}) for
this complicated case. The most standard approach would be to
perform this calculation by constructing perturbation theory in
$X_{\mu\nu}$, similar to what was done by two of us in the theory
with broken CPT or Lorentz symmetry \cite{LCPT}. The main
disadvantage of this method is that it works only for a small
$X_{\mu\nu}$. This is a reasonable assumption in the case of CPT or
Lorentz symmetry  violation, but not in the case of vector
inflation. Therefore, we shall try to make a non-perturbative in
$X_{\mu\nu}$ calculation. Later on the perturbative approach will be
used for an independent verification of the result in the linear
approximation.

In the following we consider two different ways to eliminate the
mixed $A_\mu \ph$-term in the action.

\subsection{Diagonalization by rotational type of transformation}

In order to diagonalize the operator \eq{bili} let
us make the following transformation of the fields:
\beq
&&
A_\mu = B_\mu + \frac{1}{m} \, \na_\mu \chi
\,,
\\
&&
\ph = \chi + \frac{1}{m} \, \na_\nu B^\nu
\,.
\eeq
In the new variables the action \eq{a1} becomes
\beq
\n{a2}
{\cal L}' + {\cal L}_{gf} &=&
\frac{1}{2} \, B_{\mu} \, \hat{H}_{AA} \, B_{\nu}
+ \frac{1}{2 m^2} \, \na_\mu B^\mu \, \hat{H}_{\ph\ph}
\, \na_\nu B^\nu
+ \frac{1}{m^2} \, B_{\mu}
\, X^{\mu \nu} \na_\nu \na_\al \, B^{\al}
\nonumber
\\
&+&
\frac{1}{2} \, \chi \, \hat{H}_{\ph\ph} \, \chi
+ \frac{1}{2 m^2} \, \na_\mu \chi \, \hat{H}_{AA} \, \na_\nu \chi
+ \frac{1}{m^2} \, \na_{\mu} \chi \, X^{\mu \nu} \, \na_\nu \chi
\nonumber
\\
&+&
\frac{1}{m} \, B_\mu \left[ g^{\mu\nu} (\Box - m^2)
- R^{\mu\nu} - X^{\mu\nu} \right] \na_\nu \chi
\\
&+&
\frac{1}{m} \, \chi \left[
(\Box - m^2)
+ \frac{1}{m^2} \, (\na_\mu X^{\mu\nu}) \na_\nu
+ \frac{1}{m^2} \, X^{\mu\nu} \na_\mu \na_\nu
\right] \na_\al B^\al
\nonumber
\\
&+&
\frac{1}{m} \, B_\mu \, X^{\mu\nu} \, \na_\nu \chi
+ \frac{1}{m^3}\,X^{\mu\nu} \, \na_\mu \chi \, \na_\nu \na_\al B^\al
\,.
\nonumber
\eeq
After some integration by parts one can show that all mixed terms
in Eq. \eq{a2} cancel and the Lagrangian boils down to
\beq
\n{a3}
{\cal L}' + {\cal L}_{gf} &=&
\frac{1}{2} \, B_{\mu} \, \hat{H}_{AA} \, B_{\nu}
+ \frac{1}{2 m^2} \, \na_\mu B^\mu \, \hat{H}_{\ph\ph}\,\na_\nu B^\nu
+ \frac{1}{m^2}\,B_{\mu}\,X^{\mu \nu} \na_\nu \na_\al\,B^{\al}
\nonumber
\\
&+&
\frac{1}{2} \, \chi \, \hat{H}_{\ph\ph} \, \chi
+ \frac{1}{2 m^2} \, \na_\mu \chi \, \hat{H}_{AA} \, \na_\nu \chi
+ \frac{1}{m^2} \, \na_{\mu} \chi \, X^{\mu \nu} \, \na_\nu \chi\,.
\eeq
The last expression provides a diagonal bilinear operator
$\hat{\mathbf H}$. However, there is a major problem with the
theory \eq{a3}, because of the non-minimal
$X^{\al\be}\na_\al \na_\be$-term in the vector field sector and
because of the similar non-minimal fourth derivative terms in both
scalar  and vector sectors. The operator has too complicated form,
such that no way to deal with its functional determinant is known.
For example, the generalized Schwinger-DeWitt technique of
\cite{bavi85} does not look applicable because $X^{\al\be}$
is a field and hence the non-minimal structure here can not be
parameterized by numerical parameters.

\subsection{Diagonalization by the shift-like transformation}
\n{s-try}

Consider another transformation, which involves only vector part,
\beq
&&
A^\mu = B^\mu + \al^{\mu \rho} \, \pa_\rho \ph\,,
\n{t2}
\eeq
where $B_\mu$ is the new quantum variable and $\al^{\mu \rho}$ is
an unknown function of background fields. All indices are lowered
and raised with the metric $g_{\al\be}$ and its inverse  $g^{\al\be}$.
Using the transformation (\ref{t2}) in the action \eq{a1} we arrive at
\beq
\n{a6}
{\cal L}' + {\cal L}_{gf}
&=&
\frac{1}{2} \, B_{\mu} \, \hat{H}_{AA} \, B_{\nu}
+ \frac{1}{2} \, \ph \, \hat{H}_{\ph\ph} \, \ph
+ \frac{1}{2} \, \al_{\mu \rho} \, \na^\rho \ph
\, \hat{H}_{AA} \,  \al_{\nu \om} \, \na^\om \ph
\\
&+&
\frac{1}{m} \, \al_{\nu\rho} \, \na^\rho \ph \,X^{\mu\nu}\, \na_\mu \ph
+ B_\mu \,  \left( H_{AA} \right)^{\mu\nu} \, \al_{\nu \om} \na^\om \ph
+ \frac{1}{m} \, B_{\mu} \, X^{\mu\nu} \, \na_\mu \ph\,.
\nonumber
\eeq
The condition of diagonalization is
\beq
B_\mu \Big[\,\big( H_{AA} \big)^{\mu\nu} \al_{\nu \om}
+ \frac{1}{m}  \, X^{\mu}_{\om}\,\Big]
\na^\om \ph = 0
\,,
\eeq
which has the non-local solution
\beq
\n{s1}
\al_{\mu \nu}
&=&
- \,
\frac{1}{m} \big( H_{AA} \big)^{-1}_{\mu \rho}
\,X^{\rho}_{\nu}
\nonumber
\\
&=& - \,\frac{1}{m}\,  \big(\,g^{\al\be}\Box - m^2g^{\al\be} -
R^{\al\be}-X^{\al\be}\,\big)^{-1}_{\mu \rho} \, X^{\rho}_{\nu}\,.
\eeq Let us stress that the Green function in (\ref{s1}) acts only
on the field $X^ \rho_\nu$ and not on the quantum and background
fields on the right of this expression. Replacing the solution
\eq{s1} in the Lagrangian \eq{a6} we arrive at the expression \beq
\n{a7} && {\cal L}' + {\cal L}_{gf} = \frac{1}{2} \, B_{\mu} \,
\hat{H}_{AA} \, B_{\nu} + \frac{1}{2} \, \ph \, \hat{H}_{\ph\ph} \,
\ph + \frac{1}{2m} \, \al_{\mu\rho}  \,X^{\mu\nu} \, \, \na^\rho \ph
\na_\nu \ph \,, \eeq where $\hat{H}_{AA}$ and $\hat{H}_{\ph\ph}$
were defined in Eqs.~(\ref{AA}) and (\ref{pp}), correspondingly.

The new Lagrangian, Eq. \eq{a7}, is diagonal, but it is still
non-minimal and contains a new non-local term in the scalar
field sector, with tensor $X^{\mu\nu}$ being contracted with
covariant derivatives. Once again we arrive at the situation when
the standard way of dealing with non-minimal operators
\cite{bavi85} is not applicable, because the non-minimality
can not be parameterized by a single real parameter, or even by
a finite set of such parameters.

Anyway, the result of the second approach to diagonalization has a
serious advantage compared to our first attempt. In the present case
the problem corresponds only to the scalar sector, while the form of
the vector operator is pretty standard. For instance, the expression
(\ref{a7}) is a useful starting point for making calculations
perturbatively in $\,X^{\mu\nu}$. At the same time one can do much
more than this. In the next section we describe a new efficient method
of dealing with the non-minimal scalar operator.

\section{Derivation of one-loop divergences}
\label{qua}

After applying the procedure described in  the subsection \ref{s-try},
the total action of the model consists from the pretty standard
minimal vector sector and the complicated  non-minimal scalar
action. The whole expression can be presented as follows:
\beq
\n{a-dig}
S' + S_{gf}
&=&
S_1 \,-\,
\int d^4 x \sqrt{-g}\, \left\{\frac12\, \left(g^{\mu\nu}
+ Y^{\mu\nu}\right) \pa_\mu \ph \, \pa_\nu \ph
+ \frac12\, m^2 \ph^2 \right\}\,,
\eeq
where
\beq
S_1 = \frac12\, \int d^4 x \sqrt{-g} \,
B_\mu \left( \de^\mu_\nu \Box - R^\mu_\nu - X^\mu_\nu
- \de^\mu_\nu \, m^2 \right) B^\nu
\eeq
and
\beq
Y^{\mu\nu} = \frac{1}{m^2} \, X^{\mu\nu}
- \frac{1}{m} \, \al^{\mu \rho} \, X_{\rho}^\nu\,.
\n{Ymn}
\eeq

In order to work out the scalar operator, let us define a new metric,
\beq
G^{\mu\nu}
&=&
g^{\mu\nu} + Y^{\mu\nu}\,.
\n{Gmn}
\eeq
After that the action \eq{a-dig} becomes
\beq
\n{act-G}
S' + S_{gf} &=&
S_1 \,-\,
\frac12\, \int d^4 x \sqrt{-G} \, f \,
\left( G^{\mu\nu} D_\mu \ph \, D_\nu \ph
+ m^2 \ph^2 \right)\,,
\eeq
where
\beq
f = \sqrt{\frac{\det G_{\mu \nu}}{\det g_{\mu\nu}}}
\eeq
is the new background scalar field and
$\,D_\mu \ph = \pa_\mu \ph\,$  is the covariant derivative
constructed with the new metric $\,G_{\mu\nu}$, defined as an
inverse to $\,G^{\mu\nu}$ in Eq.~ (\ref{Gmn}).  In the second
term in (\ref{act-G}) and related calculations, the indices are
lowered and raised with the new metric and with its inverse. It
proves useful to introduce the corresponding affine connection,
\beq
\Upsilon^\tau_{\al\be}
&=&
\frac12\,G^{\tau\la}\big(\pa_\al G_{\la\be}
+ \pa_\be G_{\al\la}-\pa_\la G_{\al\be}\big)\,,
\n{Up}
\eeq
covariant derivative $D_\mu$, the curvature tensor
\beq
\big[ D_\mu \,,\, D_\nu \big] \, A^{\al}
&=&
K^{\al}_{\,.\,\,\be \mu \nu} \, A^{\be}
\n{K}
\eeq
and its contractions $\,K_{\al\be} = G^{\mu\nu}\,K_{\mu\al\nu\be}\,$
and $\,K = G^{\al\be} K_{\al\be}$. These new curvatures differ from
the usual Riemann, Ricci tensors and scalar curvature $\,R\,$ by the
terms of first and higher orders in the field $X_{\mu\nu}$ and in the
non-local  $\,X_{\mu\nu}$-dependent expression $\,\al_{\mu\nu}$,
defined in Eq.~(\ref{s1}). Some useful formulas for the first order
expansions can be found in the Appendix.

Starting from this point the derivation of one-loop divergences
becomes pretty much standard. The divergences of the one-loop
effective action are given by the expression
\beq
\n{EA-G}
\Ga^{(1)}_{div} &=& \frac{i}{2}
\, \Tr \ln \hat{\mathbf H} \Big|_{div}
- i \Tr \ln \hat{H}_{gh} \Big|_{div}
\nonumber
\\
&=&
\frac{i}{2} \, \Tr \ln \left( \de^\mu_\nu \Box
- R^\mu_\nu - X^\mu_\nu
- \de^\mu_\nu \, m^2 \right)\Big|_{div}
- i \, \Tr \ln ( \Box - m^2)\Big|_{div}
\\
&+&
\frac{i}{2} \, \Tr \ln \left( D^2
+ 2 \, \hat{h}^\mu \, D_\mu - m^2 \right)\Big|_{div}\,,
\nonumber
\eeq
where
\beq
\hat{h}_\mu
&=& \frac{1}{2}\, D_\mu (\ln f)
\,=\, \frac{1}{2}\, \pa_\mu (\ln f)
\qquad
\mbox{and}
\qquad
D^2 = G^{\mu\nu} D_\mu D_\nu\,.
\n{D}
\eeq

Each of the terms in Eq.~\eq{EA-G} can be calculated separately
by means of the standard Schwinger-DeWitt technique \cite{dewitt}.
For the first term one can obtain, in dimensional regularization,
\beq
&&
\frac{i}{2} \,
\Tr \ln \left( \de^\mu_\nu \Box - R^\mu_\nu - X^\mu_\nu
- \de^\mu_\nu \, m^2 \right)\Big|_{div}
\\
&=&
- \frac{\mu^{n-4}}{\vp}\, \int d^n x \sqrt{-g}
\,\, \tr
\Big[ \frac{\hat{ 1}}{180}
\big(R_{\mu\nu\al\be}^2 - R_{\al\be}^2 + \Box R\big)
+ \frac12 \, \hat{P}^2_1 + \frac16 \, \Box \hat{P}_1
+ \frac{1}{12} \, \hat{\cal R}_{\al\be}^2\Big]\,,
\nonumber
\eeq
where $\vp = (4 \pi)^2 (n - 4)$ is the dimensional regularization
parameter, $\mu$ is the dimensional parameter of renormalization
and
\beq
\hat{1} = \de^\mu_\nu
\,, \quad
\hat{P}_1 = (P_1)^\mu_\nu =
- R^\mu_\nu + \frac{1}{6} \, \de^\mu_\nu R
- m^2 \de^\mu_\nu  - X^\mu_\nu
\,, \quad
\hat{\cal R}_{\al\be} = ({\cal R_{\al\be}})^\mu_\nu =
R^\mu_{\,.\,\nu\al\be}
\,.
\eeq
Taking into account the Faddeev-Popov ghost term, after a small
algebra we find
\beq
\n{div-BG}
&&
\frac{i}{2}\,
\Tr \ln \left( \de^\mu_\nu \Box - R^\mu_\nu - X^\mu_\nu
- \de^\mu_\nu \, m^2 \right)\Big|_{div}
- i \Tr \ln ( \Box - m^2)\Big|_{div}
\nonumber
\\
&&
=
- \frac{\mu^{n-4}}{\vp} \, \int d^4 x \sqrt{-g} \,
\left\{ - \frac{13}{180} \, R_{\al\be\mu\nu}^2
+ \frac{22}{45} \, R_{\al\be}^2
- \frac{5}{36} \, R^2
- \frac{1}{10} \, \Box R
\right.
\\
&& \left.
+\,
\frac{2}{3} \, m^2 R
+  m^4
+ m^2 X
- \frac{1}{6} \, X R
+ X_{\al\be} R^{\al\be}
+ \frac{1}{2} \, X_{\al\be}^2
- \frac{1}{6} \, \Box X
\right\}\,,
\nonumber
\eeq
where $X= g^{\al\be}X_{\al\be}$.
It is important to remember that the indices here are lowered and
raised by the normal space-time metric $\,g_{\mu\nu}$ and its
inverse.

In the scalar sector we obtain
\beq
&&
\frac{i}{2} \, \Tr \ln \left( D^2 + 2 \, \hat{h}^\mu \, D_\mu - m^2
\right)\Big|_{div}
\nonumber
\\
&&
=
- \frac{\mu^{n-4}}{\vp}\, \int d^4 x \sqrt{-G} \,
\Big[ \frac{1}{180} (K_{\mu\nu\al\be}^2 - K_{\al\be}^2 + D^2 K)
+ \frac12 \, \hat{P}^2_0 + \frac16 \, D^2 \hat{P}_0
\Big]\,,
\eeq
where
\beq
\hat{P}_0
&=&
- \,m^2 + \frac{1}{6}\, K - D_\mu  \hat{h}^\mu
- \hat{h}_\mu \, \hat{h}^\mu
\eeq
and
\beq
\frac12\, \hat{P}_0^2
&=&
\frac12\, m^4 - \frac16 \, m^2 K + \frac{1}{72}\, K^2
- m^2 F + \frac16\, K  F
+ \frac12\, F^2\,.
\eeq
In the last expression were introduced the useful notation
\beq
F
&=&
\frac{1}{4f^2}\, (D f)^2 - \frac{1}{2f} \, D^2 f
\,=\,-\,\frac{1}{\sqrt{f}}\,\big(D^2 \sqrt{f} \big)\,,
\label{F}
\eeq
where
\beq
(D f)^2
&=&
G^{\mu\nu} D_\mu f D_\nu f\,,
\qquad
 D^2 f \,=\, G^{\mu\nu} D_\mu D_\nu f\,.
\nonumber
\eeq
Thus, the unconventional scalar sector contribution can be written in the form
\beq
\n{div-s-G}
&&
\frac{i}{2} \, \Tr \ln \left( D^2 + 2 \, \hat{h}^\mu \, D_\mu - m^2
\right)\Big|_{div}
\nonumber
\\
&&
=\,
- \,\frac{\mu^{n-4}}{\vp} \, \int d^4 x \sqrt{-G} \
\left\{
 \frac{1}{180} \, K_{\al\be\mu\nu}^2
- \frac{1}{180} \, K_{\al\be}^2
+ \frac{1}{30} \, D^2 K
+ \frac{1}{72} \, K^2
\right.
\nonumber
\\
&& \left.
- \frac{1}{6} \, m^2 K
+ \frac12 \,m^4
+ \frac16 \, K F
- m^2 F
+ \frac12 \, F^2
+ \frac16\, D^2 F
\right\}\,.
\eeq

Finally, using the intermediate results \eq{EA-G}, \eq{div-BG}
and \eq{div-s-G}, we arrive at the final result for the one-loop
divergences in the theory (\ref{action}),
\beq
\Ga^{(1)}_{div}
&=& - \,\frac{\mu^{n-4}}{\vp}\, \int d^4 x \sqrt{-g} \,
\left\{
- \frac{13}{180} \, R_{\al\be\mu\nu}^2
+ \frac{22}{45} \, R_{\al\be}^2
- \frac{5}{36} \, R^2
- \frac{1}{10} \, \Box R
\right.
\nonumber
\\
&+& \left.
\frac{2}{3} \, m^2 R
+  m^4
+ m^2 X
- \frac{1}{6} \, X R
+ X_{\al\be} R^{\al\be}
+ \frac{1}{2} \, X_{\al\be}^2
- \frac{1}{6} \, \Box X
\right\}
\nonumber
\\
&-& \,\frac{\mu^{n-4}}{\vp}\, \int d^4 x \sqrt{-G} \
\left\{
 \frac{1}{180} \, K_{\al\be\mu\nu}^2
- \frac{1}{180} \, K_{\al\be}^2
+ \frac{1}{72} \, K^2
+ \frac{1}{30} \, D^2 K
\right.
\nonumber
\\
&-& \left.
\frac{1}{6} \, m^2 K
+ \frac12 \,m^4 + \frac16 \, K F - m^2 F
+ \frac12 \, F^2 + \frac16\, D^2 F
\right\}\,.
\n{EA-G-a}
\eeq

The expression (\ref{EA-G-a}) is the result of a non-standard
calculational procedure, which involves the change of quantum
variables (\ref{t2}) with the non-local coefficient defined in
(\ref{s1}). It is important to remember that the quantum field
theory calculation (Schwinger-DeWitt technique, in our case) is
applied, in the scalar sector,  to the theory with the background
metric $G^{\mu\nu}$. In terms of this new metric the second
part of the expression  (\ref{EA-G-a}) has a rather standard local
form. At the same the divergences are given by a non-local
expression in terms of the original fields $\,g_{\mu\nu}\,$ and
$\,X_{\mu\nu}$.

The Eq.~\eq{EA-G-a} enables one to obtain the one-loop
divergences in terms of the original metric $g^{\mu\nu}$ and
in each desired order in $X^{\mu\nu}$. It proves useful to
obtain an explicit expression in the first order in $\,X^{\mu\nu}$.
For this end we write down the following first-order expansions:
\beq
G_{\mu\nu}
&=&
g_{\mu\nu} - Y_{\mu\nu} + \,...
\nonumber
\\
\sqrt{- \det G^{\mu\nu}}
&=&
\sqrt{-\det g^{\mu\nu}} \Big( 1 + \frac12\, Y + \,... \Big)\,,
\eeq
where $\,Y = g^{\mu\nu} Y_{\mu\nu}$. Then for the determinant
of the inverse matrix $\,G = \det \big( G_{\mu\nu} \big)$,
\beq
\sqrt{-G}
&=&
\sqrt{-g} \Big( 1 - \frac12\, Y + ... \Big)
\nonumber
\\
f
&=&
1 + \frac12\, Y + ...
\,,\qquad
F = - \frac14 \, \Box Y + ...
\eeq
\beq
K_{\al\be\mu\nu}
&=&
R_{\al\be\mu\nu}
+ \frac12 \, \big( \na_\mu \na_\al Y_{\be\nu}
- \na_\nu \na_\al Y_{\be\mu}
+ \na_\nu \na_\be Y_{\al\mu}
- \na_\mu \na_\be Y_{\al\nu}
\nonumber
\\
&+& R^\rho_{\,.\,\al\mu\nu} \, Y_{\rho\be}
- R^\rho_{\,.\,\be\mu\nu} \, Y_{\rho\al}
\big) + ...
\nonumber
\\
K_{\al\be}
&=&
R_{\al\be}
+ \frac12\,\big(\Box Y_{\al\be} + \na_\al \na_\be Y
- \na_\rho \na_\al Y^\rho_\be - \na_\rho \na_\be Y^\rho_\al \big)
+ ...
\nonumber
\\
K
&=&
R + R_{\al\be} Y^{\al\be} + \Box Y
- \na_\al \na_\be Y^{\al\be} + ...
\eeq
\beq
D^2 = \Box + Y^{\al\be} \na_\al \na_\be
+ (\na_\al Y^{\al\be}) \na_\be
- \frac12 \, (\na^\al Y) \na_\al
+ ...
\eeq
where $D^2$ acts on a scalar field, and
\beq
Y_{\al\be} = \frac{1}{m^2} \, X_{\al\be}
+ ...
\eeq

Using these expansions we find the one-loop divergences in
the first order in $\,X^{\mu\nu}$, written in term of original
metric $g^{\mu\nu}$
\beq
\n{fin-act}
\Ga^{(1)}_{div}
&=&
\Ga^{(1)}_{vac}[g_{\mu\nu}]
\,-\, \frac{\mu^{n-4}}{\vp}\int d^4 x \sqrt{-g} \,
\Big\{
\frac34\, m^2 X
\,+\, \frac56 \, X_{\mu\nu} R^{\mu\nu}
\,-\, \frac{1}{12} \, X R
\nonumber
\\
&+&
\frac{X^{\mu\nu}}{180\, m^2} \, \Big( 5 R R_{\mu\nu}
+ 2 R^{\al\be}R_{\al\mu\be\nu}
+ 2 R_{\al\be\rho\mu} R^{\al\be\rho}_{\,.\,.\,.\,\nu}
-  4 R_{\mu\al} R^{\al}_\nu
+ 3 \Box R_{\mu\nu}
\nonumber
\\
&-&
6 \na_\mu \na_\nu R\Big)
\,-\, \frac{X}{720\,m^2} \,
\Big(2R_{\mu\nu\al\be}^2
- 2R_{\al\be}^2 + 5R^2 + 12\Box R \Big)
\Big\}\,,
\eeq
where
\beq
\Ga^{(1)}_{vac}[g_{\mu\nu}]
\,=\,
- \,\frac{\mu^{n-4}}{\vp} \int d^4 x \sqrt{-g} \,
\Big\{
\frac{29}{60} \, R_{\al\be}^2
- \frac{1}{15} \, R_{\al\be\mu\nu}^2
- \frac{1}{8} \, R^2
+ \frac{m^2}{2} \,  R
+ \frac{3m^4}{2} \,
\Big\}
\n{Proca-div}
\eeq
is the contribution of Proca field, which depends only on external
metric and not on $X^{\mu\nu}$.  In case of $\,X_{\mu\nu}=0$ we come
back to the well-known result derived in \cite{bavi85,fervi,BuGui}
by means of different  methods. In the formula \eq{fin-act} we omitted
the total derivative terms for the sake of brevity. As one should expect,
the linear result is local in terms of the original fields
$\,g_{\mu\nu}\,$ and $\,X_{\mu\nu}$.


\section{Calculation using universal traces}
\label{pert}

As we have already mentioned above, the main expression
(\ref{EA-G-a}) is the result of a non-standard calculational
procedure, hence it would be useful to have its verification.
The first order in $X_{\mu\nu}$ divergences can be obtained
independently by means of the universal functional traces
method (generalized Schwinger-DeWitt technique) of Barvinsky
and Vilkovisky \cite{bavi85}.  In this section we shall perform
such a calculation in order to compare the result with
Eq.~\eq{fin-act}, and thus partially verify the main result
(\ref{EA-G-a}). The calculation described below is analogous
to the one developed in Ref.~\cite{LCPT} for the complicated
gauge model with broken Lorentz and CPT background. For
this reason we can skip some technical explanations, which
can be found in this reference.

Before the diagonalization procedure, the bilinear
form \eq{bili}, can be written as
\beq
\hat{\mathbf H} = \hat{\mathbf H}_m + \hat{\mathbf H}_{nm}
\,,
\eeq
where
\beq
\hat{\mathbf H}_m = \hat{\bf 1} \Box
+ 2 \, \hat{\bf L}^{\mu} \, \na_\mu
+ (\hat{\bf \Pi}_0 + \hat{\bf M})
\,
\eeq
is the minimal part of bilinear operator in quantum fields and
\beq
\hat{\mathbf H}_{nm} = \hat{\bf K}^{\mu\nu} \na_\mu \na_\nu
\eeq
is the non-minimal part. The relevant matrices are defined as follows:
\beq
\hat{\bf 1}
&=&
\left(\begin{array}{cc}
\de^\mu_\nu & 0 \\
0 & 1
\end{array} \right) \,,
\nonumber
\\
\hat{\bf L}^\mu
&=&
\left(\begin{array}{cc}
0 & \frac{1}{2m} \, X^{\mu\nu} \\
- \frac{1}{2m} \, X^{\mu\nu} & \frac{1}{2m^2} \, \na_\nu X^{\mu\nu}
\end{array} \right) \,,
\nonumber
\\
\hat{\bf \Pi}_0
&=&
 \left(\begin{array}{cc}
- \de^\mu_\nu \,m^2 - R^\mu_\nu & 0 \\
0 & -m^2
\end{array} \right) \,,
\nonumber
\\
\hat{\bf M}
&=&
\left(\begin{array}{cc}
- X^\mu_\nu  & 0 \\
- \frac{1}{m} \, \na_\nu X^{\mu\nu} & 0
\end{array} \right)\,,
\nonumber
\\
\hat{\bf K}^{\mu\nu}
&=&
\left(\begin{array}{cc}
0  & 0 \\
0 & \frac{1}{m^2} \, X^{\mu\nu}
\end{array} \right) \,.
\eeq

The one-loop effective action is given by the known formula
\beq
\n{EA-G-BV}
\Ga^{(1)} &=& \frac{i}{2} \, \Tr \ln (\hat{\mathbf H}_m
+ \hat{\mathbf H}_{nm})
- i \, \Tr \ln \hat{H}_{gh}
\,.
\eeq
Let us make the following transformation:
\beq
\n{exp}
\Tr \ln \hat{\mathbf H} &=&
\Tr \ln ( \hat{\mathbf H}_m + \hat{\mathbf H}_{nm})=
\Tr \ln \hat{\mathbf H}_m + \Tr \ln ( \hat{\bf 1} + \hat{\mathbf H}_m^{-1}
\,.\, \hat{\mathbf H}_{nm})
\nonumber
\\
&=&
\Tr \ln \hat{\mathbf H}_m
+ \Tr \hat{H}_{nm} \,.\, \hat{H}_0^{-1}
+ \dots\,,
\eeq
where dots stand for the terms of higher orders in $X^{\mu\nu}$,
$\,\hat{H}_0 = \Box - m^2\,$ and
\beq
\n{h0i}
\hat{H}_{nm}
&=&
\frac{1}{m^2} \, X^{\mu\nu} \na_\mu \na_\nu \,.
\eeq
In the last line of Eq.~(\ref{exp}) we perform the expansion of the
logarithm and take into account only terms of the first order in
$X^{\mu\nu}$. The first term in the last line of Eq.~\eq{exp} can
be directly calculated by the standard Schwinger-DeWitt method
\cite{dewitt}, while the second term can be calculated by means
of the universal functional traces method \cite{bavi85}.

For the minimal part it is possible to obtain the one-loop
divergences by using the known formula of the Schwinger-DeWitt
technique,
\beq
\n{div-min}
\frac{i}{2} \Tr \ln \hat{\mathbf H}_m \Big|_{div}
\,=\,
- \frac{\mu^{n-4}}{\vp} \int d^4 x \sqrt{-g}  \tr
\Big[ \frac{\hat{\bf 1}}{180} (R_{\mu\nu\al\be}^2 - R_{\al\be}^2)
+ \frac12 \hat{\bf P}^2 + \frac{1}{12} \hat{\bf S}_{\al\be}^2
\Big],
\eeq
where
\beq
\hat{\bf P} = \hat{\bf P}_0
+ \hat{ \bf M} - \na_\mu \hat{\bf L}^\mu + ...\,,
\eeq
\beq
\hat{\bf S}_{\al\be}
= {\cal \hat{\bf R}}_{\al\be}
- \na_\al \hat{\bf L}_\be + \na_\be \hat{\bf L}_\al + ...\,,
\eeq
with
\beq
\hat{\bf P}_0 = \hat{\bf \Pi}_0 + \frac16 \, \hat{\bf 1} \, R
\qquad
\mbox{and}
\qquad
{\cal \hat{\bf R}}_{\al\be} = \hat{\bf 1} \, [\na_\al,\na_\be]
\,.
\eeq
Up to the first order in the new parameters, we obtain
\beq
\frac12 \, \tr \hat{\bf P}^2
&=&
 \frac12 \, \tr \hat{\bf P}_0^2
+ \tr \hat{\bf P}_0 \, \hat{\bf M}
- \tr \hat{\bf P}_0 \, \na_\mu \hat{\bf L^\mu} + ...\,,
\nonumber
\\
\tr \hat{\bf S}_{\al\be}^2
&=&
\tr ( {\cal \hat{\bf R}}_{\al\be}^2
+ 4 \, {\cal \hat{\bf R}}^{\al\be} \, \na_\be \hat{\bf L}_\al )
+ \dots\,.
\eeq
while Eq.~\eq{div-min} with the Faddeev-Popov ghost term
reduce to
\beq
&&
\frac{i}{2} \, \Tr \ln \hat{\mathbf H}_m \Big|_{div}
- i \, \Tr \ln \hat{H}_{gh} \Big|_{div}
\,=\,
\Ga^{(1)}_{vac}[g_{\mu\nu}]
\nonumber
\\
&&
-\,\frac{\mu^{n-4}}{\vp} \,\int d^4 x \sqrt{-g}\,
\tr \Big[ \hat{\bf P}_0 \,\hat{\bf M}
- \hat{\bf P}_0 \, \na_\mu \hat{\bf L}^\mu
\,+\,
\frac13 \,{\cal \hat{\bf R}^{\al\be}}\,\na_\be \hat{\bf L}_\al \Big]
\,+\dots\,,
\eeq
where $\Ga^{(1)}_{vac}[g_{\mu\nu}]$ was defined in
Eq.~\eq{Proca-div}. After performing some algebra one can find
\beq
\tr \hat{\mathbf P}_0 \,\hat{\mathbf M}
&=&
m^2 X + X_{\al\be} R^{\al\be}  - \frac16 \,X R + ...\,,
\nonumber
\\
\tr \hat{\mathbf P}_0 \, \na_\mu \hat{\mathbf L}^\mu
&=&
\frac{1}{12 m^2} \, X^{\al\be} \na_\al \na_\be R + ...\,,
\nonumber
\\
\tr {\cal \hat{\bf R}^{\al\be}} \,\na_\be \hat{\bf L}_\al
&=&
0 + ...\,.
\eeq
So, finally
\beq
\n{div-m}
\hspace{-0.8cm}
\frac{i}{2} \, \Tr \ln \hat{\mathbf H}_m \Big|_{div}
&-&
i \, \Tr \ln \hat{H}_{gh} \Big|_{div}
\,=\,
- \frac{\mu^{n-4}}{\vp}\, \int d^4 x \sqrt{-g} \,
\Big[  m^2 X + X_{\al\be} R^{\al\be}
\nonumber
\\
&-& \frac{1}{12 m^2} \, X^{\al\be} \na_\al \na_\be R
 - \frac16 \,X R
 \Big]
\,+\, \Ga^{(1)}_{vac}[g_{\mu\nu}] \,+\, ...\,.
\eeq

For calculating the divergent part of the non-minimal piece of
Eq.~\eq{exp} we need the inverse of the operator $\hat{H}_0$,
which can be expressed as
\beq
\hat{H}_0^{-1} = \frac{1}{\Box} + m^2 \, \frac{1}{\Box^2}
+ m^4 \, \frac{1}{\Box^3} + {\cal O} (l^{-5})\,.
\eeq
In the last formula $\,{1/\Box}\,$ is the inverse of d'Alembert
operator and the last term $\,{\cal O}(l^{-5})\,$ indicates omitted
irrelevant terms of a higher background dimension \cite{bavi85}.

Using equation \eq{h0i} one can obtain the relation
\beq
\n{nmH0}
\Tr \hat{H}_{nm} \,.\, \hat{H}_0^{-1}
&=&
X^{\mu\nu}
\Big(
\frac{1}{m^2}\, \na_\mu \na_\nu \, \frac{1}{\Box}
+ \na_\mu \na_\nu  \, \frac{1}{\Box^2}
+ m^2 \na_\mu \na_\nu \, \frac{1}{\Box^3}\Big)
\,+\, {\cal O} (l^{-5})\,.
\eeq
The equation \eq{nmH0} is already in the form that allows us to
apply the tables of universal functional traces of  \cite{bavi85}.
The calculation is straightforward (albeit not really easy) and
the results have the form (note that we use Minkowski signature)
\beq
\frac{1}{m^2} \, X^{\mu\nu} \na_\mu \na_\nu \,
\frac{1}{\Box} \Big|_{div}
&=&
\frac{i\mu^{n-4}}{\vp}\,
\int d^4 x \sqrt{-g} \, \Big\{
\frac{1}{m^2}\,X^{\mu\nu}\,
\Big(\frac{1}{45} R_{\al\be\rho\mu}\, R^{\al\be\rho}_{\,.\,.\,.\,\nu}
+ \frac{1}{45} R^{\al\be}\, R_{\al\mu\be\nu}
\nonumber
\\
&-&
\frac{2}{45} R_{\mu\al} \, R^{\al}_\nu
+ \frac{1}{18 } \, R \, R_{\mu\nu}
+ \frac{1}{30}\Box R_{\mu\nu}
+ \frac{1}{10} \na_\mu \na_\nu R\Big)
\nonumber
\\
&-&
\frac{1}{m^2} \, X \Big( \frac{1}{180} \, R_{\mu\nu\al\be}^2
- \frac{1}{180} \, R_{\al\be}^2 + \frac{1}{72} \, R^2
+ \frac{1}{30} \, \Box R \Big)\Big\}\,,
\nonumber
\\
X^{\mu\nu} \na_\mu \na_\nu  \, \frac{1}{\Box^2} \Big|_{div}
&=&
\frac{i\mu^{n-4}}{\vp}
\int d^4 x \sqrt{-g} \,
\Big\{ \frac16 \, X R - \frac13\,X_{\al\be} R^{\al\be}
\Big\}\,,
\nonumber
\\
\n{tf}
m^2 \, X^{\mu\nu} \na_\mu \na_\nu  \, \frac{1}{\Box^2} \Big|_{div}
&=&
- \,\frac{i\mu^{n-4}}{2\vp}\, \int d^4 x \sqrt{-g} \, m^2 X \,,
\eeq

By using relations \eq{nmH0} and \eq{tf} one can obtain
\beq
\n{div-nm}
\frac{i}{2}\, \Tr \hat{H}_{nm} \,.\, \hat{H}_0^{-1} \Big|_{div}
&=&
- \frac{\mu^{n-4}}{\vp} \int d^4 x \sqrt{-g} \,
\Big\{
\frac{1}{12} \, X R
- \frac16 \, X_{\mu\nu} R^{\mu\nu}
- \frac14 \, m^2 X
\nonumber
\\
&+&
\frac{1}{m^2} \, X^{\mu\nu}
\Big(
\frac{1}{90}\,R_{\al\be\rho\mu}\,R^{\al\be\rho}_{\,.\,.\,.\,\nu}
+ \frac{1}{90} \, R^{\al\be}\, R_{\al\mu\be\nu}
- \frac{1}{45} \, R_{\mu\al} \, R^{\al}_\nu
\nonumber
\\
&+&
\frac{1}{36} \, R \, R_{\mu\nu}
+ \frac{1}{60} \,\Box R_{\mu\nu}
+ \frac{1}{20} \, \na_\mu \na_\nu R
\Big)
\nonumber
\\
&-&
\frac{1}{2m^2} \, X \, \Big( \frac{1}{180} \, R_{\mu\nu\al\be}^2
- \frac{1}{180} \, R_{\al\be}^2 + \frac{1}{72} \, R^2
+ \frac{1}{30} \, \Box R \Big)
\Big\}\,.
\eeq
Now, from equations \eq{EA-G-BV}, \eq{exp}, \eq{div-m}
and \eq{div-nm} it is not difficult to verify that we arrive exactly
at the result for the one-loop divergences derived by the new method, Eq.~\eq{fin-act}.

\section{Conclusions and discussions}
\label{con}

Let us summarize the results. We have developed the generic
procedure for consistent formulation of the effective action for the
the free massive Abelian vector in curved space-time with
non-minimal coupling to an arbitrary symmetric background tensor
$X_{\mu\nu}$. The result of the calculation for the divergent part
of effective action has the form (\ref{EA-G-a}). The remarkable
feature of this expression is that it has a standard appearance of a
local expression when written in terms of the quantities $K$, $D$
and $F$, and at the same time it looks as a rather unusual non-local
functional when expressed in terms of the original external field
$g_{\mu\nu}$ and $X_{\mu\nu}$. The reason for this is the non-local
change of variable which has been performed in the course of
diagonalization of the bilinear form of the action. Nevertheless,
this effective action can be expanded in functional power series in
$X_{\mu\nu}$. Then we will obtain infinite number of local terms
containing any powers of $X_{\mu\nu}$ with any number of covariant
derivatives action on $X_{\mu\nu}$. Number of derivatives will
increase with power of expansion.

It would be quite interesting to explore some extensions and
modifications of this result. First of all it would be interesting
to formulate a similar scheme for constructing the effective action
and calculation of one-loop divergences for the massless version of
the theory, which has an important applications to astrophysics
\cite{WT}. The development of this model would open the way for
exploring the self-interacting massless vector field, with the
possibility to analyze the dynamical symmetry breaking in the vector
model.

An obvious possible extension of our model is related to introducing
the interaction to fermions and other components of the Standard
Model of Particle Physics. Such an extension would be especially
useful for the applications to inflation, because this would opens
the gate to explore particle creation on the vector inflation.

On the other hand, it would be certainly interesting to add a
self-interaction term, but this is not a simple task. For instance,
the most natural form $(A_\mu A^\mu )^ 2$ may be quite
complicated to deal with,
as it was discussed in \cite{book} in relation to the model of
conformal invariant axial vector (antisymmetric torsion) field.
The theory with such a term, without  $X^{\mu\nu}$ and mass,
possess local conformal invariance, has soft breaking of gauge
invariance and was never explored at both classical and quantum
levels. The massive version looks more accessible and will be
hopefully explored in the next publications. We leave this problem
for the future, expecting that the technical progress achieved in
the present work would be useful for this end.

\section*{Acknowledgements}
The study of I.L.B was supported by The Ministry of Education and
Science of Russian Federation, project 3.1386.2017. Also he is
grateful to the RFBR grant, project No 15.02.03594. T.P.N. wish to
acknowledges CAPES for the support through the PNPD fellowship.
I.Sh. is grateful to CNPq, FAPEMIG and ICTP for partial support of
his work.

\section*{Appendix. First-order relations for
curvatures}

Here we present the useful expansion formulas for the terms
in Eq.~\eq{fin-act}, which are of the fourth order in the
extended covariant derivative $\,D_\mu$,
\beq
K_{\al\be\mu\nu}^2
&=&
R_{\al\be\mu\nu}^2 + 2  \,  R_{\al\be\rho\mu} \,
R^{\al\be\rho}_{\,.\,.\,.\,\nu} \, Y^{\mu\nu}
+ 4 \, R^{\al\be\mu\nu} \, \na_\mu \na_\al Y_{\be\nu}
\,+\, ...
\nonumber
\\
&=&
R_{\al\be\mu\nu}^2
+ 4 \, Y^{\al\be} \, \Box R_{\al\be} - 2 \, Y ^{\al\be} \, \na_\al \na_\be R
+ 2   \, R_{\al\be\rho\mu} \, R^{\al\be\rho}_{\,.\,.\,.\,\nu} \, Y^{\mu\nu}
\nonumber
\\
&+& 4 \, R^{\al\be} \, R_{\al\mu\be\nu} \, Y^{\mu\nu}
- 4 \, R_{\al\be} \, R^{\be}_\mu \, Y^{\mu\al}
\,+\, \mbox{total derivatives} \,,
\eeq
\beq
K_{\al\be}^2
&=&
R_{\al\be}^2 + R_{\al\be}\big(
2R^{\be}_\mu \, Y^{\mu\al}
+ \Box Y^{\al\be}
+ \na^\al \na^\be Y
- 2 \na_\rho \na^\al Y^{\rho\be}\Big)
\,+\, ...
\nonumber
\\
&=&
R_{\al\be}^2
+ Y^{\al\be} \, \Box R_{\al\be}
+ \frac12 \, Y \, \Box R
- Y^{\al\be} \, \na_\al \na_\be R
+ 2 \, R^{\al\be} \, R_{\al\mu\be\nu} \, Y^{\mu\nu}
\nonumber
\\
&+& \mbox{total derivatives} \,,
\eeq
\beq
K^2
&=&
R^2
+ 2R R_{\al\be}\, Y^{\al\be}
+ 2 Y \Box R
-  2 Y^{\al\be}\, \na_\al \na_\be R
\,+\, \mbox{total derivatives} \,,
\eeq
\beq
D^2 K
&=&
Y^{\al\be} \na_\al \na_\be R
+ (\na_\al Y^{\al\be}) \na_\be R
- \frac12 \, (\na^\al Y) \na_\al R
\,+\, \mbox{total derivatives}
\nonumber
\\
&=& \frac12 \, Y \Box R
\,+\, \mbox{total derivatives} \,.
\eeq




\begin{thebibliography}{m}

\bibitem{PP}
E. Allys, J.P. Beltran Almeida, P. Peter, and Y. Rodr\'{\i}guez,
{\it On the 4D generalized Proca action for an Abelian vector field},
JCAP {\bf 1609} (2016) 026,
arXiv:1605.08355.

\bibitem{Dim1}
K.~Dimopoulos, and M.~Karciauskas,
{\it Non-minimally coupled vector curvaton},
JHEP {\bf 0807} (2008) 119,
arXiv:0803.3041; 
\\
K.~Dimopoulos, 
{\it Statistical Anisotropy and the Vector Curvaton Paradigm,}
Int. J. Mod. Phys. {\bf  D21} (2012) 1250023, 
Erratum: {\it ibid,} 1292003, 
arXiv:1107.2779.

\bibitem{Tasinato} 	
G. Tasinato,
{it Cosmic Acceleration from Abelian Symmetry Breaking},
JHEP {\bf 1404} (2014) 067,
arXiv:1402.6450.

\bibitem{Heis-2014}  L. Heisenberg,
{\it  Generalization of the Proca Action},
JCAP {\bf 1405} (2014) 015,
\\
arXiv:1402.7026.

\bibitem{DeRham-2014} C. De Rham, L. Keltner, and A.J. Tolley,
{\it  Generalized galileon duality},
Phys. Rev. {\bf D90} (2014) 024050,
arXiv:1403.3690.

\bibitem{HeisKimYam} L. Heisenberg, R. Kimura, and K. Yamamoto,
{\it  Cosmology of the proxy theory to massive gravity},
Phys. Rev. {\bf D89} (2014) 103008,
arXiv:1403.2049.

\bibitem{Farid}
F. Charmchi, Z. Haghani, S. Shahidi, and L. Shahkarami,
{\it  One-loop corrections to vector Galileon theory},
Phys. Rev. {\bf D93} (2016) 124044,
arXiv:1511.07034;
\\
A. Amado, Z. Haghani, A. Mohammadi, and Sh. Shahidi,
{\it Quantum corrections to the generalized Proca theory via
a matter field},
arXiv:1612.06938.

\bibitem{LH-mn}	
J.B. Jimenez, and L. Heisenberg,
{\it Derivative self-interactions for a massive vector field},
Phys. Lett. {\bf B757} (2016) 405, 
arXiv:1602.03410;
\\	
A. De Felice, L. Heisenberg, R. Kase, Sh. Mukohyama, Sh. Tsujikawa,
and Y. Zhang,
{\it Cosmology in generalized Proca theories},
JCAP {\bf 1606} (2016) 048,
arXiv:1603.05806;
\\
{\it Effective gravitational couplings for cosmological
perturbations in generalized Proca theories},
Phys.Rev. {\bf D94} (2016) 044024,
arXiv:1605.05066;
\\
L. Heisenberg, R. Kase, Sh. Tsujikawa,
{\it Beyond generalized Proca theories },
Phys.Lett. {\bf B760} (2016) 617, 
DOI: 10.1016/j.physletb.2016.07.052\
arXiv:1605.05565;
\\
J.B. Jimenez, L. Heisenberg,
{\it Generalized multi-Proca fields},
arXiv:1610.08960.

\bibitem{WT} M.S. Turner, and L.M. Widrow,
{\it  Inflation Produced, Large Scale Magnetic Fields},
Phys. Rev. {\bf D37} (1988) 2743.

\bibitem{vec-in} A. Golovnev, V. Mukhanov,
and V. Vanchurin,
{\it   Vector Inflation},
JCAP {\bf 0806} (2008) 009.

\bibitem{Jabbari}
A. Maleknejad, M.M. Sheikh-Jabbari, and J. Soda,
{\it  Gauge Fields and Inflation},
Phys. Rept. {\bf 528} (2013) 161,   
arXiv:1212.2921.

\bibitem{Uzan} G. Esposito-Farese, C. Pitrou, and J.-Ph. Uzan,
{\it  Vector theories in cosmology},
Phys. Rev. {\bf D81} (2010) 063519,
arXiv:0912.0481

\bibitem{book}
I.L. Buchbinder, S.D. Odintsov, and I.L. Shapiro,
{\it Effective Action in Quantum Gravity}
(IOP Publishing, Bristol, 1992).

\bibitem{Stueckelberg} H. Ruegg, and M. Ruiz-Altaba,
{\it The Stueckelberg field},
Int. J. Mod. Phys. {\bf A19} (2004) 3265, 
hep-th/0304245.

\bibitem{Toms} D.J. Toms,
{\it   Quantization of the minimal and non-minimal vector
field in curved space,}
arXiv:1509.05989.

\bibitem{FJ} L. Faddeev, and R. Jackiw,
{\it Hamiltonian Reduction of Unconstrained and Constrained Systems},
Phys. Rev. Lett. {\bf 60} (1988) 1692.

\bibitem{dewitt} B.S.~DeWitt,
{\it Dynamical theory of groups and fields}
(Gordon and Breach, New York, 1965).

\bibitem{bavi85} A.O.~Barvinsky, and G.A.~Vilkovisky,
{\it The Generalized Schwinger-DeWitt Technique in Gauge
Theories and Quantum Gravity}, Phys. Repts. {\bf 119} (1985) 1.

\bibitem{fervi} E.V. Gorbar, and I.L. Shapiro,
{\it Renormalization Group and Decoupling in Curved Space:
II. The Standard Model and Beyond,}
JHEP {\bf 06} (2003) 004, hep-ph/0303124.

\bibitem{BuGui}
I.L. Buchbinder, G. de Berredo-Peixoto, and I.L. Shapiro,
{\it  Quantum Effects in Softly Broken Gauge Theories
in Curved Space-Times},
Phys. Lett. {\bf B649} (2007) 454, 
 hep-th/0703189.

\bibitem{BF} A. Belokogne, and A. Folacci,
{\it Stueckelberg massive electromagnetism in curved spacetime:
Hadamard renormalization of the stress-energy tensor and the
Casimir effect},
Phys.Rev. {\bf D93} (2016) 044063,
arXiv:1512.06326.

\bibitem{LCPT}
T.~de P.~Netto and I.L.~Shapiro,
{\it Vacuum contribution of photons in the theory with
Lorentz and CPT-violating terms},
Phys. Rev. {\bf D89} (2014) 104037, arXiv:1402.3152.

\end{thebibliography}
\end{document}